\documentclass[aps,preprint,tightenlines,showpacs,nofootinbib]{revtex4}

\usepackage{graphicx}  
\usepackage{bm}  
\usepackage{amsmath}
\usepackage{epsf}
\newcommand{\beq}{\begin{equation}}
\newcommand{\eeq}{\end{equation}}
\newcommand{\bea}{\begin{eqnarray}}
\newcommand{\eea}{\end{eqnarray}}

\begin{document}
\title{Novel evaluation of the two-pion contribution 
       to the nucleon isovector form factors}
\author{M.A. Belushkin}\email{belushki@itkp.uni-bonn.de}
\author{H.-W. Hammer}\email{hammer@itkp.uni-bonn.de}
\affiliation{Helmholtz-Institut f\"ur Strahlen- und Kernphysik (Theorie),
Universit\"at Bonn, Nu\ss allee 14-16, D-53115 Bonn, Germany}

\author{Ulf-G. Mei\ss ner}\email{meissner@itkp.uni-bonn.de}
\affiliation{Helmholtz-Institut f\"ur Strahlen- und Kernphysik (Theorie),
Universit\"at Bonn, Nu\ss allee 14-16, D-53115 Bonn, Germany\\
and\\
Institut f\"ur Kernphysik (Theorie), Forschungszentrum
J\"ulich, D-52425 J\"ulich, Germany\\}

\date{\today}
\begin{abstract}
We calculate the two-pion continuum contribution to the nucleon
isovector spectral functions drawing upon the new high statistics
measurements of the pion form factor by the CMD-2, KLOE, and SND 
collaborations. 
The general structure of the spectral functions remains unchanged,
but the magnitude increases by about 10\%.
Using the updated spectral functions, we calculate the contribution 
of the two-pion continuum to the nucleon isovector form factors and
radii. We compare the isovector
radii with simple $\rho$-pole models and illustrate
their strong underestimation in such approaches. Moreover,
we give a convenient parametrization of the result for use in future form 
factor analyses. 
\end{abstract}
\pacs{11.55.Fv, 13.40.Gp, 14.20.Dh}
\maketitle
\section{Introduction}

The electromagnetic form factors of the nucleon offer a unique window
on strong interaction dynamics over a wide range of momentum
transfers \cite{Gao:2003ag,Hyde-Wright:2004gh}.
At small momentum transfers, one is only sensitive
to the gross properties of the nucleon like the charge and magnetic moment,
while at high momentum transfers one will be able to resolve aspects
of the quark substructure of the nucleon as described by QCD.
The form factors encode both perturbative and non-perturbative aspects
of QCD such as the nucleon radii, vector meson coupling constants, 
and the asymptotic behavior at large momentum transfer.
Moreover, their detailed understanding is important
for a wide variety of experiments ranging from
the strange vector form factors of the nucleon \cite{strange} to
Lamb shift measurements in atomic hydrogen \cite{Lamb}.

It has been known for a long time that the pion plays an important role in 
the long--range structure of the nucleon \cite{FHK}. This connection
was made more precise using dispersion theory in the 1950's
\cite{CKGZ58,FGT58}. Subsequently, Frazer and Fulco have written down 
partial wave dispersion relations that relate the nucleon electromagnetic 
structure to pion-nucleon ($\pi N$) scattering and predicted the existence of
the $\rho$ resonance \cite{FF,FF60a}.
Despite of this success, the central role of the $2\pi$ continuum
in the isovector spectral function
has often been ignored in vector-meson dominance analyses of the 
electromagnetic form factors of the nucleon where the $\rho$ was 
approximated by a simple pole. In 1975,
H\"ohler and Pietarinen pointed out that this omission leads to
a gross underestimate of the isovector radii of the nucleon \cite{HP2}. 
H\"ohler et al. first perfomed a consistent 
dispersion analysis of the electromagnetic form factors of the
nucleon \cite{Hohler:1976ax} including the $2\pi$ continuum
derived from the pion form factor and $\pi N$-scattering data \cite{HP}.
In the mid-nineties, this analysis has been updated \cite{MMD} and extended
to include data in the time-like region \cite{HMD}. Recently, the new precise 
data for the neutron electric form factor have been included as 
well \cite{Hammer:2003ai}.

Using chiral perturbation theory (CHPT), the long-range pionic structure
of the nucleon can be connected to the Goldstone boson dynamics of QCD
\cite{BKMrev}. The nonresonant part of
the $2\pi$ continuum is in excellent agreement with the
phenomenological analysis \cite{BKMspec} and the $\rho$-meson
contribution can be included as well \cite{KM,KM2,Norbert}.
It is well known that vector mesons play an important 
role in the electromagnetic structure of the nucleon, see e.g. 
\cite{FF,Sak,GoSa,HP2,Gari,Dubni,Lomo}, and the remaining contributions
to the spectral function
have usually been approximated by vector meson resonances. 

A new twist to this picture was recently given in 
Ref.~\cite{FW}, where the form factor data were interpreted based 
on a phenomenological fit with an ansatz for the pion cloud using the 
old idea that the proton can be thought of as virtual neutron-positively 
charged pion pair. A very long-range contribution to the charge distribution 
in the Breit frame extending out to about 2~fm was found and attributed to 
the pion cloud. While naively the pion Compton wave length is of this size, 
these findings are indeed surprising if compared with the ``pion cloud'' 
contribution due to the $2\pi$ contribution for the isovector spectral 
functions of the nucleon form factors, which can be obtained from unitarity or 
chiral perturbation theory. As was shown in Ref.~\cite{Hammer:2003qv},
these latter contributions to  the long-range part of the nucleon 
structure are much more confined in coordinate space 
and agree well with earlier (but less systematic) calculations based on chiral 
soliton models, see e.g. \cite{UGM}. Therefore it remains to be shown how to 
reconcile the findings of Ref.~\cite{FW}, based on a global fit to all nucleon 
form factors, with the results of dispersion analysis and chiral perturbation 
theory.

The CMD-2 \cite{CMD2}, KLOE \cite{KLOE}, and SND \cite{SND}
collaborations have recently remeasured the pion form factor
with high statistics and found significant deviations from earlier 
measurements at the $\rho$-resonance peak. Moreover, the three
measurements also show discrepancies among each other.
In light of the importance of the 
$2\pi$ continuum for the electromagnetic structure of the nucleon,
it appears worthwhile to recalculate the $2\pi$-continuum contribution
to the isovector spectral function of the nucleon using the new
high statistics data and estimate the errors that arise from the 
discrepancy among the experiments.
Moreover, a new analysis will help to better understand the nature and
range of the pion cloud of the nucleon.

\section{Nucleon Form Factors}

First, we collect some basic definitions. The nucleon 
electromagnetic (EM) form factors are defined by the nucleon matrix element 
of the electromagnetic current, 
\beq 
\langle N(p') | J_\mu^{\rm EM} | N(p)\rangle 
= \bar u (p') \left[ \gamma^\mu \, F_1 (t) + \frac{i}{2m} 
\sigma^{\mu\nu} (p'-p)_\nu \,  F_2 (t) \right] u(p)~, 
\eeq 
with $t= (p'-p)^2 =q^2<0$ the invariant momentum transfer squared, 
and $m$ the nucleon mass. $F_1 (t)$ 
and $ F_2 (t)$ are the Dirac and the Pauli form factors, respectively. 
They are normalized at $t=0$ to the charge ($F_1$) and 
anomalous magnetic moment ($F_2$). 
It is convenient to work in the isospin basis and to
decompose the nucleon form factors into isoscalar ($S$) and isovector 
($V$) parts,
\begin{equation}
F_i^S = \frac{1}{2} (F_i^p + F_i^n) \, , \quad
F_i^V = \frac{1}{2} (F_i^p - F_i^n) \, ,\quad i=1,2\,,
\end{equation}
where $p$ ($n$) denotes the proton (neutron).
The experimental data are usually given for the Sachs form factors,
which are linear combinations of $F_1$ and $F_2$
\begin{equation}
\label{sachs}
G_{E}^I(t) = F_1^I(t) - \tau F_2^I(t) \, , \quad
G_{M}^I(t) = F_1^I(t) + F_2^I(t) \, , \quad I = S, V\,, 
\end{equation}
where $\tau = -t/(4 m^2)$.
In the Breit frame, where the energy transfer of the virtual photon vanishes,
$G_{E}$ and $G_{M}$ may be interpreted as the Fourier transforms of the 
charge and magnetization distributions in coordinate space,
respectively.

The analysis of the  nucleon 
electromagnetic form factors proceeds most
directly through the spectral representation given by\footnote{We work here 
with unsubtracted dispersion relations. Since the normalizations of all the 
form factors are known, one could also work with once-subtracted dispersion 
relations. For the topic studied here, this is of no relevance.} 
\beq\label{disprel} 
F_i^I (t) = \frac{1}{\pi}\, \int_{t_0^I}^\infty 
\frac{{\rm Im}\,F_i^I (t')\, dt'}{t' - t}~, \quad i =1,2\,, 
\quad I = S, V\,, 
\eeq 
where the corresponding thresholds are given by 
$t_0^S = (3M_\pi)^2$ and $t_0^V = (2M_\pi)^2$, respectively and
$M_\pi$ is the pion mass.
The imaginary part entering Eq.~(\ref{disprel})
can be obtained from a spectral decomposition \cite{FGT58}.
For this purpose it is convenient to consider the
EM current matrix element in the time-like region,
\beq
\label{eqJ}
J_\mu = \langle N(p) \overline{N}(\bar{p}) | J_\mu^{\rm em} | 0 \rangle
= \bar{u}(p) \left[ F_1 (t) \gamma_\mu +\frac{i}{2 m} \sigma_{\mu\nu}
(p+\bar{p})^\nu F_2 (t) \right] v(\bar{p})\,,
\eeq
where $p\,,\bar{p}$ are the momenta of the nucleon-antinucleon pair
created by the current $J_\mu^{\rm em}$. The four-momentum transfer
in the time-like region is $t=(p+\bar{p})^2 > 0$.
Using the LSZ reduction formalism, the imaginary part
of the form factors is obtained by inserting a complete set of
intermediate states as \cite{FGT58}
\beq
\label{spectro}
{\rm Im}\,J_\mu = \frac{\pi}{Z}(2\pi)^{3/2}{\cal N}\,\sum_\lambda
 \langle p | \bar{J}_N (0) | \lambda \rangle 
 \langle \lambda | J_\mu^{\rm em}| 0 \rangle \,v(\bar{p})
\,\delta^4(p+\bar{p}-p_\lambda)\,,
\eeq
where ${\cal N}$ is a nucleon spinor normalization factor, $Z$ is
the nucleon wave function renormalization, and $\bar{J}_N (x) =
J_N^\dagger(x) \gamma_0$ with $J_N(x)$ a nucleon source.
The states $|\lambda\rangle$ are asymptotic states of
momentum $p_\lambda$. They carry the same quantum numbers as
the current $J^{\rm em}_\mu$: $I^G(J^{PC})=0^-(1^{--})$ for
the isoscalar component and $I^G(J^{PC})=1^+(1^{--})$ for the
isovector component of $J^{\rm em}_\mu$.
Furthermore, they have no net baryon number.

For the isoscalar current  the lowest mass states are: $3\pi$,
$5\pi$, $\ldots$; for the isovector part they
are: $2\pi$, $4\pi$, $\ldots$. Because of $G$-parity, states
with an odd number of pions only contribute to the isoscalar
part, while states with an even number contribute to the
isovector part. 
Associated with each intermediate state is a
cut starting at the corresponding threshold in $t_0$ and running to
infinity. As a consequence,
the spectral function ${\rm Im}\, F(t)$ is different from zero along the
cut from $t_0$ to $\infty$ with $t_0 = 4 \, (9) \, M_\pi^2$ for the
isovector (isoscalar) case.
Using Eqs.~(\ref{eqJ},\ref{spectro}), the spectral functions for
the form factors can in principle be constructed from experimental
data. In practice, this program can only be carried out for the 
lowest-mass two-particle intermediate states ($2\pi$ and $K\bar{K}$)
\cite{HP,Hammer:1998rz,Hammer:1999uf}.

The longest-range (and therefore most important at low momentum
transfer) pion cloud contribution comes from the $2\pi$ intermediate state 
in the  isovector form factors. This contribution was first 
constructed from the pion form factor and $\pi N$ scattering data 
by H\"ohler and Pietarinen \cite{HP}. 

\section{Two--Pion Continuum}

In this paper, we re-evaluate the $2\pi$ contribution in a 
model--independent way using the latest experimental data for the pion
form factor from CMD-2 \cite{CMD2}, KLOE \cite{KLOE}, and
SND \cite{SND}.
We follow  Ref.~\cite{LB} and express the $2\pi$ contribution to the
the isovector spectral functions  in terms of the pion 
charge form factor $F_\pi (t)$ and the P--wave $\pi\pi \to \bar N N$ 
amplitudes $f^1_\pm(t)$. The $2\pi$ continuum is expected to be the
dominant contribution to the isovector spectral function 
from threshold up to masses of about 1~GeV \cite{LB}.
Here, we use the expressions
\bea 
\nonumber 
{\rm Im}~G_E^{V} (t) &=& \frac{q_t^3}{m\sqrt{t}}\, 
F_\pi (t)^* \, f^1_+ (t)~,\\ 
{\rm Im}~G_M^{V} (t) &=& \frac{q_t^3}{\sqrt{2t}}\, 
F_\pi (t)^* \, f^1_- (t)~, 
\label{uni} 
\eea 
where $q_t=\sqrt{t/4-M_\pi^2}$. The imaginary parts of the Dirac
and Pauli Form factors can be obtained using Eq.~(\ref{sachs}).
The  P--wave $\pi\pi \to \bar N N$ amplitudes $f_\pm^1(t)$ are tabulated in  
Ref.~\cite{LB}. (See also Ref.~\cite{Pieta} for an unpublished
update that is consistent with Ref.~\cite{LB}.) 
We stress that the representation of Eq.~(\ref{uni}) 
gives the exact isovector spectral functions for $4M_\pi^2 \leq t 
\leq 16 M_\pi^2$, but in practice holds up to $t \simeq 50 M_\pi^2$.
Since the contributions from $4\pi$ and higher 
intermediate states is small up to  $t \simeq 50 M_\pi^2$, $F_\pi(t)$
and the $f_\pm^1(t)$ share the same phase in this region and the 
two quantities can be replaced by their absolute values.\footnote{
We note that representation of Eq.~(\ref{uni}) is most useful for
our purpose. The manifestly real functions $J_\pm (t) = f_\pm^1(t)/ 
F_\pi (t)$ also tabulated in Ref.~\cite{LB}
contain assumptions about the pion form factor
which leads to inconsistencies when used together with the 
updated $F_\pi (t)$.}

\begin{figure}[t] 
\centerline{\includegraphics*[width=12cm,angle=0]{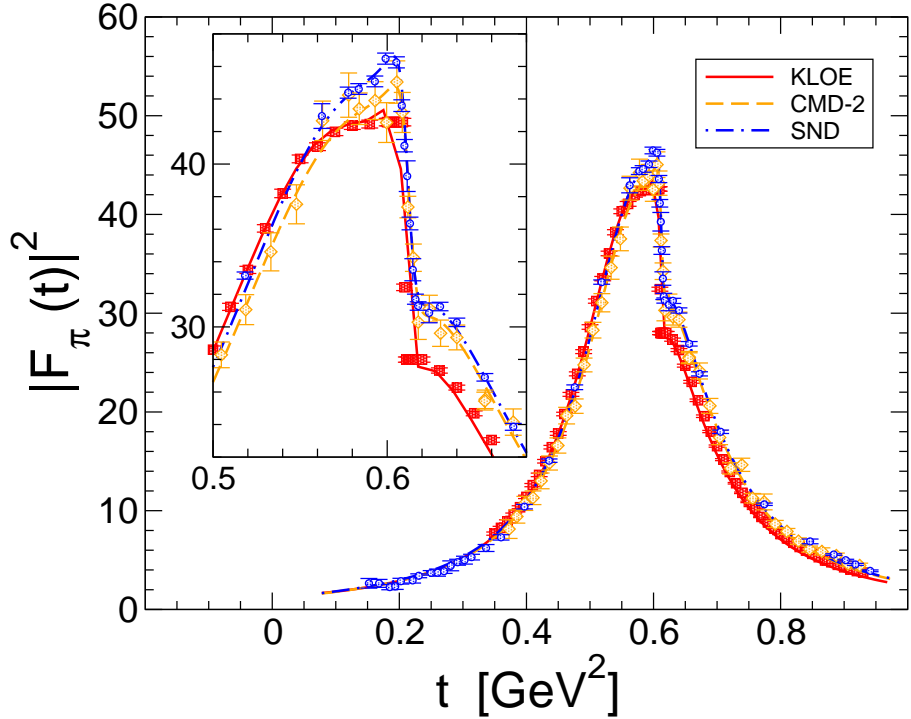}}
\caption{
The pion electromagnetic form factor in the time-like
region as a function of the momentum transfer $t$. The diamonds,
squares, and circles show the high statistics data from the 
CMD-2 \cite{CMD2}, KLOE \cite{KLOE}, and SND \cite{SND}
collaborations, respectively. The dashed, solid, and dash-dotted 
lines are our model parametrizations. The inset shows
the discrepancy in the resonance region in more detail.
}
\label{piFF}
\end{figure}

The updated pion form factor is shown in Fig.~\ref{piFF}.
The diamonds, squares, and circles show the high statistics data from the 
CMD-2 \cite{CMD2}, KLOE \cite{KLOE}, and SND \cite{SND}
collaborations, respectively. The dashed, solid, and dash-dotted 
lines are our model parametrizations which are of the Gounaris-Sakurai type
\cite{MMD,GoSa}. The form factor shows a pronounced 
$\rho$-$\omega$ mixing in the vicinity of the $\rho$-peak.
There are discrepancies between the three experimental data sets for the 
pion form factor \cite{SND}. The discrepancies in the 
$\rho$-resonance region are  shown in more detail in the inset
of Fig.~\ref{piFF}. Since we are not in the 
position to settle this experimental problem, we will take the
three data sets at face value. We will evaluate the  $2\pi$ continuum 
given by Eq.~(\ref{uni})
for all three sets and estimate the errors from the discrepancy 
between the sets.

\begin{figure}[ht] 
\centerline{\includegraphics*[width=12cm,angle=0]{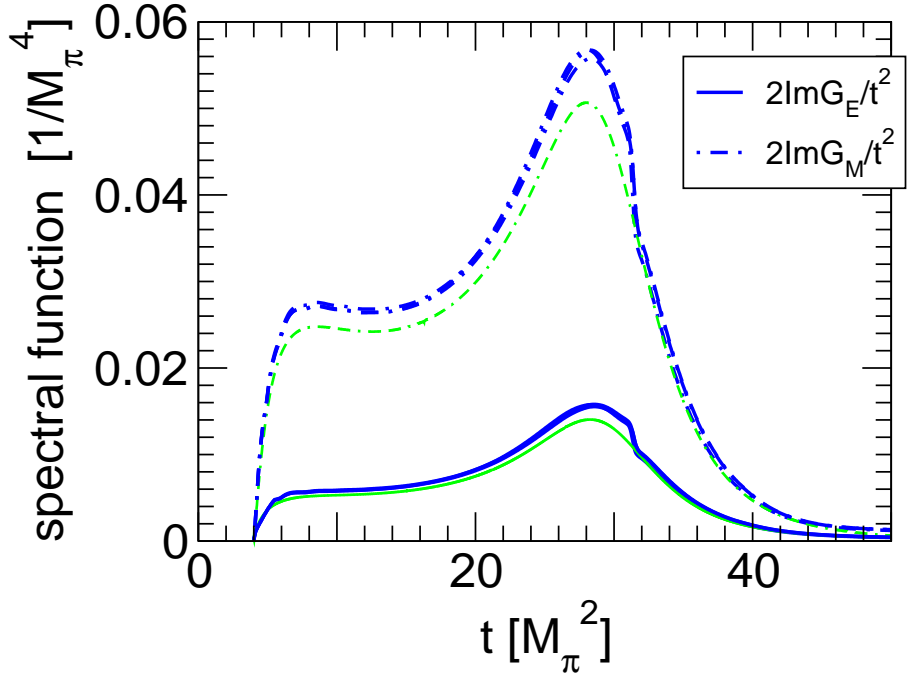}}
\caption{
The $2\pi$ spectral function using the new high statistics data for 
the pion form factor \cite{CMD2,KLOE,SND}. The spectral functions weighted by 
$1/t^2$ are shown for $G_E$ (solid line) and $G_M$ (dash-dotted line). 
The previous results by H\"ohler et al.~\cite{LB} (without $\rho$-$\omega$ 
mixing) are shown for comparison by the gray/green lines.}
\label{2pispec}
\end{figure} 
Using the new high statistics pion form factor data \cite{CMD2,KLOE,SND}
and the amplitudes $f_\pm^1(t)$
tabulated in  Ref.~\cite{LB}, we obtain the spectral functions
shown in Fig.~\ref{2pispec}.
We show the spectral functions weighted by 
$1/t^2$ $G_E$ (solid line) and $G_M$ (dash-dotted line). The previous 
results by H\"ohler et al.~\cite{LB} (without $\rho$-$\omega$ 
mixing) are given for comparison  by the gray/green lines.
The general structure of the two
evaluations is the same, but there is a difference in magnitude of about 10\%.
The difference between the three data sets for the pion form factor is
very small and indicated by the line thickness. The difference in the
form factors is largest in the $\rho$-peak region (cf.~Fig.~\ref{piFF}),
but this region is suppressed by the $\pi\pi \to \bar N N$ amplitudes 
$f_\pm^1(t)$ which show a strong fall-off as $t$ increases.

The spectral functions have two distinct features. First, as already 
pointed out in \cite{FF}, they contain the important contribution of 
the $\rho$-meson with its peak at $t \simeq 30 M_\pi^2$. 
Second, on the left shoulder of the $\rho$, the isovector spectral functions 
display a very pronounced enhancement close to the two-pion threshold. This 
is due to the logarithmic singularity on the second Riemann sheet located at 
$t_c = 4M_\pi^2 - M_\pi^4/m^2 = 3.98 M_\pi^2$, very close to the threshold. 
This pole comes from the projection of the nucleon Born graphs, or in modern 
language, from the triangle diagram. If 
one were to neglect this important unitarity correction, one would severely 
underestimate the nucleon isovector radii \cite{HP2}. In fact, precisely the 
same effect is obtained at leading one-loop accuracy  in 
relativistic chiral perturbation 
theory \cite{GSS,UGMlec}. This topic was further elaborated on in 
the framework of heavy baryon CHPT \cite{BKMspec,Norbert} and in a covariant 
calculation based on infrared regularization \cite{KM}. Thus, the most 
important $2\pi$ contribution to the nucleon form factors can be determined 
by using either unitarity or CHPT (in the latter case, of course, the $\rho$ 
contribution is not included).

\begin{table}[htb]
\begin{tabular}{|l|c|c|}
\hline
Ref. & \quad $\langle r^2\rangle^V_1$ [fm$^2$] \quad & 
\quad $\langle r^2\rangle^V_2$ [fm$^2$]\quad \\
\hline
this work & 0.32...0.33 & 1.77...1.80  \\
Ref.~\cite{Norbert} & 0.19 & 1.10 \\
Ref.~\cite{iacho} $(i)$ & 0.19 & 0.96 \\
Ref.~\cite{iacho} $(ii)$ & 0.27 & 1.38 \\
\hline
\end{tabular}
\caption{
Comparison of the $2\pi$-continuum contribution to the 
Dirac and Pauli isovector radii of the nucleon (first line) to 
three different $\rho$-pole models from Refs.~\cite{Norbert,iacho}. 
The given range indicates the error introduced by the different
pion form factor data sets.
}
\label{tab:radii}
\end{table}
The importance of the correct $2\pi$-continuum contribution for the 
nucleon isovector radii is illustrated in Table \ref{tab:radii}. We compare 
the contribution of the full $2\pi$ continuum and various $\rho$-pole
parametrizations to the Dirac and Pauli isovector radii of the 
nucleon \cite{HP2}:
\beq
\langle r^2\rangle^V_i = \frac{6}{\pi} \int_{4 M_\pi^2}^{50 M_\pi^2}
\frac{{\rm Im}\,F_i^V (t)}{t^2}dt\,, \quad i =1,2\,. 
\eeq
The first line shows the
contribution of the full $2\pi$ continuum from this work to the
nucleon isovector radius. The given range indicates the error introduced by 
the different pion form factor data sets.
The second line shows the $\rho$-pole
parametrization used in Ref.~\cite{Norbert}. The third and fourth lines show
the $\rho$-pole contribution from Ref.~\cite{iacho} excluding $(i)$
and including $(ii)$ an approximate $2\pi$ continuum, respectively. 
It is obvious that the simple $\rho$-pole parametrizations from lines
two and three underestimate the full contribution by about 30-40\%
depending on the form factor. The approximate $2\pi$ continuum
from  Ref.~\cite{iacho} in the fourth line comes fairly close for 
the Dirac form factor but still underestimates the Pauli radius
by about 20\%.

Inserting the spectral functions into the dispersion relation
Eq.~(\ref{disprel}), we obtain the $2\pi$ contribution to the 
nucleon isovector form factors. The contribution of the spectral
function in the region $t \geq 50 M_\pi^2$ is very small and set to 
zero in this evaluation.
The results for the form factors can be fitted by an
expression of the form \cite{Hohler:1976ax,MMD}
\beq
\label{FFvpara}
F_i^V(t) = \frac{a_i +b_i (1-t/c_i)^{-2/i}}{2(1-t/d_i)}\,,\quad i=1,2\,,
\eeq
where $a_i$, $b_i$, $c_i$, and $d_i$ are constants.
Averaging the results for the three different pion form factor
data sets \cite{CMD2,KLOE,SND}, the values of the constants are
$a_1=1.10788$, $b_1=0.109364$, $c_1=0.36963$, $d_1=0.553034$,
$a_2=5.724253$, $b_2=1.111128$, $c_2=0.27175$, 
and $d_2=0.611258$.
The errors in these constants are of the order 4\% or less.
Using the parametrization from Eq.~(\ref{FFvpara}), the $2\pi$ contribution 
to the isovector form factors can
easily be included in any form factor analysis. It is fixed by 
the pion form factor and $\pi N$ data and contains no free parameters.

\section{Summary \& Conclusion}
In this letter, we have presented a novel analysis of the $2\pi$
contribution to the nucleon isovector spectral function 
using the new high statistics data
of the pion form factor by the CMD-2 \cite{CMD2},
KLOE \cite{KLOE}, and SND \cite{SND} collaborations.
The difference in the spectral function
between the three data sets for the pion form factor is very small. 
The spectral function displays the contribution of the $\rho$ peak around 
$t \simeq 30 M_\pi^2$ and the pronounced enhancement close to the two-pion 
threshold at $t = 4M_\pi^2$ from the logarithmic singularity on the 
second Riemann sheet.
The magnitude of the spectral function increases by about 10\% compared to
the previous analyses.

The conclusions of Ref.~\cite{Hammer:2003qv} about the 
possibility of a long-range pion cloud remain unaffected by this
change in magnitude: a long-range
pion cloud extending as far as 2 fm is not compatible with what is known
about the $2\pi$ contribution to the nucleon isovector spectral
function. We note that this conclusion might have to be modified 
if the higher-mass pion-continua ($3\pi$, $4\pi$,...) show a
significant threshold enhancement similar to the $2\pi$ continuum.
Given the current state of knowledge, however, this appears unlikely.
In Ref.~\cite{BKMspec}, the threshold behavior of the  $3\pi$ continuum 
was explicitly calculated in heavy baryon ChPT and no enhancement was found. 
Moreover, the  inelasticity from four pions in $\pi\pi$ scattering 
and four-pion production in $e^+ e^-$ annihilation at low momentum
transfer are known to be small \cite{LB,Gasser:1990bv,Ecker:2002cw}.

Finally, we have calculated the resulting contribution to the 
nucleon isovector form factors and given a convenient parametrization of 
the result. This contribution is fixed from the pion form factor and
$\pi N$-scattering data and contains no free parameters. It can easily 
be used as an independent input in future form factor analyses. 
This will reduce the number of free parameters and ensure the correct 
spectral function on the nearest part of the cut in the time-like region. 
A new dispersion-theoretical analysis of the nucleon form factors 
using the updated $2\pi$ continuum is in preparation \cite{new}.

\begin{acknowledgments}
We thank A. Denig for providing the KLOE data for the pion form factor
and the referee for pointing out the CMD-2 and SND measurements.
This work was supported
in part by the EU Integrated Infrastructure Initiative Hadron Physics
under contract number RII3-CT-2004-506078
and the Deutsche Forschungsgemeinschaft through funds provided
to the SFB/TR 16 \lq\lq Subnuclear structure of matter''.
\end{acknowledgments}

 
\end{document}